\documentclass[aps, prd, showpacs, nofootinbib, preprintnumbers]{revtex4}
\usepackage{graphicx,color}
\usepackage{amsmath,amssymb}
\newcommand{\VEV}[1]{\langle #1 \rangle}
\newcommand{\tr}{\mathrm{tr}}

\newcommand{\diag}{\mathrm{diag}}
\newcommand{\gBL}{g_{B-L\rule{0mm}{7pt}}}
\newcommand{\gBLsq}{g^2_{B-L\rule{0mm}{6pt}}}
\newcommand{\gBLquad}{g^4_{B-L\rule{0mm}{6pt}}}
\newcommand{\gtld}{g_{\mathrm{mix}}}
\newcommand{\gtldsq}{g^2_{\mathrm{mix}}}

\begin{document}
\title{Radiative Symmetry Breaking from Flat Potential 
in various $U(1)'$ models}
\preprint{KEK-TH 1700}
\preprint{OU-HET 805}
\author{Michio Hashimoto}
\email{michioh@isc.chubu.ac.jp}
\affiliation{
Chubu University,  
Kasugai-shi,  Aichi, 487-8501, JAPAN }

\author{Satoshi Iso}
\email{satoshi.iso@kek.jp}
\affiliation{
Theory Center, KEK and
Sokendai,  
Tsukuba, Ibaraki 305-0801, JAPAN }

\author{Yuta Orikasa}
\email{orikasa@het.phys.sci.osaka-u.ac.jp}
\affiliation{
Department of Physics, 
Osaka University,  Toyonaka, Osaka 560-0043, JAPAN}




\date{\today}

\begin{abstract}

We investigate a radiative electroweak gauge symmetry breaking scenario
via the Coleman-Weinberg mechanism starting from a completely flat 
Higgs potential at the Planck scale (``flatland scenario''). 
In our previous paper, we showed that the flatland scenario is possible 
only when an inequality $K<1$ among the coefficients of 
the $\beta$ functions is satisfied.
In this paper, we calculate the number $K$ in various models 
with an extra $U(1)$ gauge sector in addition to the SM particles. 
We also show the renormalization group (RG) behaviors of 
a couple of the models as examples.
\end{abstract}

\maketitle

\section{Introduction}

Recently the ATLAS and CMS groups reported a discovery of 
a new particle like the Higgs boson in the standard model (SM) and 
the mass is near 125~GeV~\cite{Higgs-discover}.
This means that within the framework of the SM, the perturbation works up to
the Planck scale $M_{Pl}$. 
It is well-known, however, 
that this value of mass causes the so-called stability problem of the SM vacuum.
Compared with the vacuum expectation value (VEV) of the Higgs field,
$v=246$~GeV, the Higgs mass is relatively small and thus 
the Higgs potential encounters instability,
if we calculate the renormalization group (RG) improved effective potential 
with the running Higgs quartic coupling $\lambda_H(\mu)$ up to 
$\mu = M_{Pl}$, where $\mu$ is the renormalization scale~\cite{stability}. 
If the instability occurs at  $\mu < M_{Pl}$, it may 
suggest an appearance of new physics below the Planck scale $M_{Pl}$.
On the other hand, if the Higgs potential is stable up to $M_{Pl}$ and 
vanishes there, an interesting possibility can be indicated that 
the Higgs is borne at the Planck scale as a scalar field 
with a flat potential~\cite{Froggatt:1995rt,Shaposhnikov:2009pv,Holthausen:2011aa}.

Another problem associated with the Higgs potential 
is that the Higgs mass receives enormous radiative 
corrections by, if they exist, heavy particles coupled to the Higgs boson. 
Related to this large radiative corrections to the Higgs mass squared,
the naturalness problem has been vigorously examined.
Admittedly, the supersymmetry in the TeV scale resolves 
the naturalness problem, but the LHC and other experiments have given
strong constraints on their masses both directly and indirectly. 
Similarly, the Technicolor scenario was a beautiful idea, 
but it has been faced with several difficulties,
the large $S$-parameter, the smallness of the discovered Higgs mass, etc.

Because of these difficulties, alternative solutions to 
the naturalness problem are widely discussed~\cite{naturalness} these days. 
Suppose that the UV completion theory (which may be beyond the ordinary
field theories like the string theory) is connected with the SM sector
in a way that the SM has no dimensionful parameters. 
Then if no large intermediate mass scales exist between the SM and 
the UV completion theory,
large logarithmic corrections violating 
the multiplicative renormalization of the Higgs mass term are never 
generated: the SM becomes free from the naturalness problem.
Such models based on the above idea are called classically conformal 
models with no intermediate scales~\cite{ccm,Iso:2009ss,Iso:2009nw,Iso:2012jn}. 
In these models, the classical Lagrangian contains no mass terms and all 
dimensionful parameters are dynamically generated. 

Motivated by the vacuum stability and the naturalness problem, 
we proposed a model that 
the electroweak (EW) symmetry is radiatively broken 
in the infrared (IR) region via the Coleman-Weinberg mechanism (CWM)~\cite{CW} 
starting from a flat scalar potential in the ultraviolet (UV) 
region~\cite{Iso:2012jn}.
A more radical possibility that all the scalar potentials vanish 
at the UV scale was proposed in Ref.~\cite{Chun:2013soa}
and the RG flows were numerically calculated. 
(See also the corrigendum to \cite{Chun:2013soa}.)
It is, however, rather nontrivial to construct such a model because 
various couplings must be finely tuned so that
the running scalar quartic coupling vanishes
both in the IR and the UV regions. 
In Ref.~\cite{Hashimoto:2013hta}, we gave a general criterion 
for such a scenario, which we call a ``flatland scenario''. 
We showed that an inequality $K < 1$ must be satisfied
among the coefficients of the $\beta$ functions 
in order to realize the flatland. 

In our previous paper~\cite{Hashimoto:2013hta},
we considered the $B-L$ model~\cite{Langacker:2008yv}  
and calculated the number $K$.
In this paper, we investigate a wider class of the $U(1)$ extension 
models classified in Ref.~\cite{Carena:2004xs}. 
Although the Higgs doublets are assumed to have no charge of
the extra $U(1)$ in Ref.~\cite{Carena:2004xs}, 
we may assign the extra $U(1)$ charges to the Higgs doublets
so as to write the SM Yukawa couplings in terms of the mass-dimension 
four operators.
Because there appears the $Z$--$Z'$ mixing in this case, 
the $\rho$-parameter deviates from unity at the tree level.
We give a constraint on the model parameters from it.
We also propose the minimal vector-like $U(1)'$ model for 
the flatland scenario, where the SM fermions do not have 
the $U(1)'$ charges and the spontaneous symmetry breaking (SSB)
in the CW sector is transferred into the electroweak symmetry breaking (EWSB) 
in the SM sector by vector-like fermions having both charges of 
the SM and $U(1)'$.
This vector-like model can be connected with 
semi-invisible $Z'$ models widely discussed in the literature~\cite{dark-Z}. 

The paper is organized as follows:
In Sec.~\ref{sec2}, we study what kind of the $U(1)$ extension models
can satisfy the necessary condition for the flatland scenario.
In Sec.~\ref{sec3}, we discuss the $\rho$-parameter.
In Sec.~\ref{sec4}, we show realizations of the flatland scenario.
Sec.~\ref{summary} is devoted to summary and discussions.
The full set of the renormalization group equations (RGEs) are
shown in Appendix~\ref{app-u1p} and \ref{app-vec}. 

\section{$U(1)'$ models and Coleman-Weinberg mechanism}
\label{sec2}

It is well known that the CWM does not work within the SM
because of the large top Yukawa coupling.
Thus we need to extend the SM by introducing an additional
sector in which the dynamical mass generation occurs.

We study several extensions of 
the $U(1)$ sector in the SM including the right-handed neutrinos 
$\nu_R$~\cite{Carena:2004xs}.
We also introduce the minimal vector-like $U(1)'$ model.

Let us consider a scenario that the Higgs potential is totally flat 
at some high energy scale $\Lambda$
and that the extra $U(1)$ breaking via the CWM is encoded into 
the EWSB through the quartic scalar mixing term.
This flatland scenario can be realized only when a certain 
necessary condition is satisfied~\cite{Hashimoto:2013hta}.
Below we discuss whether or not the flatland scenario works
in the $U(1)$ extension models.

It is convenient to rescale the extra $U(1)$ gauge coupling $g_{1'}$
so as to assign the charge of $\nu_R$ to $-1$.
The Majorana Yukawa coupling is then 
\begin{equation}
  {\cal L}_M = - Y_M^{ij} \overline{\nu_{Ri}^c} \, \nu_{Rj} \Phi
  + \mbox{(h.c.)} ,
  \label{yM}
\end{equation}
where the charge of the SM singlet Higgs $\Phi$ is $+2$.
The VEV of the extra scalar $\Phi$ breaks the $U(1)'$ gauge symmetry and 
provides the mass of $Z'$.
For simplicity,
we may take $Y_M^{ij} = \diag (y_M,\cdots,y_M, 0,\cdots,0)$ and
$\tr [(Y_M^{ij})^{2}] = N_\nu y_M^{2}$, etc.,
where $N_\nu$ stands for the number of the right-handed neutrinos 
having relevant Majorana couplings.
The SM Yukawa couplings are discussed separately in a model-dependent way.
The scalar potential for $\Phi$ is
\begin{equation}
  V = \lambda_\Phi |\Phi|^4 + \cdots, 
\end{equation}
where the last $(\cdots)$ terms are model-dependent.
We assume that the Higgs potential has classical conformality,
i.e., all of the scalar mass squared terms are vanishing.

The relevant RGEs for the flatland scenario 
are written as~\cite{Hashimoto:2013hta} 
\begin{eqnarray}
  \beta_{g_{1'}} &\equiv& \mu \frac{\partial }{\partial \mu} g_{1'} =
  \frac{a}{16 \pi^2}  g_{1'}^3, \label{rge-g} \\
  \beta_{y_M} &\equiv& \mu \frac{\partial }{\partial \mu} y_M =
  \frac{y_M}{16 \pi^2} \bigg[\, b y_M^2 - c g_{1'}^2\,\bigg], \label{rge-yM} \\
  \beta_{\lambda_\Phi} &\equiv& 
  \mu \frac{\partial }{\partial \mu} \lambda_\Phi =
  \frac{1}{16 \pi^2} \bigg[\, - d y_M^4 + f g_{1'}^4 + \cdots \,\bigg],
  \label{rge-lam}
\end{eqnarray}
with~\cite{Basso:2010jm}
\begin{equation}
  a = \frac{2}{3} \sum_{f} Q_f Q_f + \frac{1}{3} \sum_{s} Q_s Q_s,
  \label{coeff-a}
\end{equation}
where $Q_f$ and $Q_s$ are the charges of the extra $U(1)$ for
two-component (chiral) fermions and complex scalars, respectively.
The last dots in $\beta_{\lambda_\Phi}$ include
$\lambda_\Phi^2$, $\lambda_\Phi g_{1'}^2$, etc., which are irrelevant 
in the following analysis.

In order to realize the flatland scenario, 
the $\beta$ function for $\lambda_\Phi$ should satisfy
$\beta_{\lambda_\Phi} > 0$ in the IR region and simultaneously
$\beta_{\lambda_\Phi} < 0$ in the UV region.
(For example, see Fig.1 in Ref.~\cite{Hashimoto:2013hta}.)
Owing to the IR fixed point of the RGEs~\cite{Pendleton:1980as}, 
we can encode the above inequalities of $\beta_{\lambda_\Phi}$
into the following necessary condition~\cite{Hashimoto:2013hta},  
\begin{equation}
K \equiv \frac{a+c}{b} \sqrt{\frac{d}{f}} < 1,
\label{nec-cond}
\end{equation}
which is written only in terms of the coefficients of the $\beta$ functions.
Unless the inequality is satisfied, 
without any concrete calculations for the RG flows, we can conclude that
the CWM does not work starting from the flat potential at the UV scale. 
In this sense, the necessary condition $K < 1$ is a powerful tool
for the model-building.
Because we rescaled the extra $U(1)$ charge to impose $Q_{\nu_R} = -1$,
we obtain $b$, $c$, $d$, $f$, model-independently,
\begin{equation}
  b = 4+2N_\nu, \quad c = 6, \quad 
  d=16N_\nu, \quad f=96 ,
\end{equation}
where we assumed that only the right-handed neutrinos are coupled 
with the SM singlet scalar $\Phi$.
Even if some other fermions interact with $\Phi$, we obtain 
similar results depending on the form of the interactions.
The remaining labor is now only to calculate 
the coefficient $a$ of $\beta_{g_{1'}}$ in each $U(1)$ extension.

In the following subsections, we will study whether or not
the necessary condition $K < 1$ can be satisfied
in several chiral $U(1)$ extensions~\cite{Carena:2004xs}. 
We also introduce the minimal vector-like $U(1)'$ model with $K < 1$.
\subsection{$U(1)_{xq-\tau^3_R}$}

\begin{table}
 \begin{center}
$$
  \begin{array}{c|ccc|c} \hline
    & SU(3)_c & SU(2)_W & U(1)_Y & U(1)_{xq-\tau^3_R} \\ \hline
   q_L & {\bf 3} & {\bf 2} & \frac{1}{6} & x \\
   u_R & {\bf 3} & {\bf 1} & \frac{2}{3} & 4x-1 \\
   d_R & {\bf 3} & {\bf 1} & - \frac{1}{3} & -2x+1 \\ \hline
   \ell_L & {\bf 1} & {\bf 2} & - \frac{1}{2} & -3x \\
   \nu_R & {\bf 1} & {\bf 1} & 0 & -1 \\
   e_R & {\bf 1} & {\bf 1} & -1 & -6x+1 \\ \hline
   H & {\bf 1} & {\bf 2} & + \frac{1}{2} & 3x-1 \\
   \Phi & {\bf 1} & {\bf 1} & 0 & +2 \\ \hline
  \end{array}
$$
  \end{center}
  \caption{Charge assignment for $U(1)_{xq - \tau^3_R}$. 
   The $x$-charges with $x=1/3$ and $x=0$ correspond to 
   the $B-L$ and $U(1)_R$ models, respectively. \label{charge1}}
\end{table}

Let us consider the $U(1)$ extension shown in Table~\ref{charge1}.
In Ref.~\cite{Carena:2004xs}, this is represented by $U(1)_{q+xu}$.
Notice that we rescaled the $U(1)$ charges so as to impose 
the charge of $\nu_R$ to $-1$, as in the $B-L$ model.
We may call it $U(1)_{xq - \tau^3_R}$, whose charges correspond to 
$6xY - \tau_R^3$. 
Taking $x=0$ or $x=1/3$, 
this model corresponds to $U(1)_R$ or $U(1)_{B-L}$ model, respectively.

The Higgs doublet $H$ does not have the extra $U(1)$ charge 
in Ref.~\cite{Carena:2004xs}, 
in order to avoid the $\rho$-parameter constraint. 
In this case, there is no contribution to the coefficient $a$ 
of $\beta_{g_{1'}}$ from $H$, while the SM Yukawa coupling is not 
the conventional one.
Instead, we may assign the extra $U(1)$ charge to $H$ so as to
get the conventional Yukawa couplings\footnote{
This is essentially the same as the model discussed in 
the corrigendum to \cite{Chun:2013soa}.}.
One of the cost is that the $\rho$-parameter deviates from unity
at the tree level. We will discuss it later.

The SM Yukawa couplings are then
\begin{equation}
 -{\cal L}_y = y_u \bar{q}_L u_R \tilde{H} + y_d \bar{q}_L d_R H
  + y_\ell \bar{\ell}_L e_R H + y_\nu \bar{\ell}_L \nu_R \tilde{H} ,
\end{equation}
where $\tilde{H} \equiv i\tau_2 H^*$.
The coefficient $a$ of $\beta_{g_{1'}}$ is
\begin{equation}
  a = \left( 80 x^2 -32 x + \frac{16}{3} \right) \, N_g 
    + \frac{4}{3} N_\Phi + \frac{2}{3} (3x-1)^2 N_H ,
\end{equation}
where $N_g$ represents the number of generations\footnote{
One may regard them as extra (or ``hidden'') generations like in
Ref.~\cite{Lee:2012xn}.} 
which couple to the extra $U(1)$ gauge field, 
$N_\Phi \; (=1)$ is the number of $\Phi$,
and $N_H \; (=1)$ denotes the number of $H$.
We then find
\begin{equation}
  K = \frac{\left( 40 x^2 - 16 x + \frac{8}{3} \right) \, N_g 
    + \frac{2}{3} N_\Phi + \frac{1}{3} (3x-1)^2 N_H + 3}{2+N_\nu}
    \sqrt{\frac{N_\nu}{6}} \, .
\end{equation}
We show the full set of the RGEs in Appendix~\ref{app-u1p}.

For the $B-L$ extension, i.e., $x=1/3$, we obtain\cite{Hashimoto:2013hta} 
\begin{equation}
  K = \frac{\frac{16}{9} N_g + \frac{2}{3} N_\Phi + 3}{2+N_\nu}
    \sqrt{\frac{N_\nu}{6}} \, .
\end{equation}
For $N_g=1,2$ and $N_\nu=1$, we find
$K=0.74, 0.98$, respectively, and otherwise, $K > 1$~\cite{Hashimoto:2013hta}.
Unfortunately, the familiar $B-L$ model with $N_g=3$ does not work.

How about the general case with an arbitrary $x$? 

For $N_g=3$ and $N_\Phi=N_H=1$, the minimum of $a$ is 
$a_{\rm min} = 964/123$, when $x_{\rm min}=25/123$. 
We then find the minimum values of $K$ as follows:
\begin{equation}
 K_{\rm min} = 0.9415, \quad 0.9986, \quad 0.9785 \quad
 \mbox{for} \quad N_\nu=1,2,3, 
\end{equation}
respectively.
Even for $N_g=3$, the flatland scenario can work.

We will discuss later the realization of the flatland scenario in 
this minimal case.
Also, the constraint of the $\rho$-parameter is studied.
\subsection{$U(1)_{10+x\bar{5} + yY}$}

\begin{table}
 \begin{center}
$$
  \begin{array}{c|ccc|c} \hline
    & SU(3)_c & SU(2)_W & U(1)_Y & U(1)_{10+x\bar{5}+yY} \\ \hline
   q_L & {\bf 3} & {\bf 2} & \frac{1}{6} & x + y\\
   u_R & {\bf 3} & {\bf 1} & \frac{2}{3} & -x + 4y \\
   d_R & {\bf 3} & {\bf 1} & - \frac{1}{3} & 1 - 2x -2y \\ \hline
   \ell_L & {\bf 1} & {\bf 2} & - \frac{1}{2} & 2x -3y -1 \\
   \nu_R & {\bf 1} & {\bf 1} & 0 & -1 \\
   e_R & {\bf 1} & {\bf 1} & -1 & -x -6y\\ \hline
   \nu'_R & {\bf 1} & {\bf 1} & 0 & 1-5x \\ \hline
   \psi_L^\ell & {\bf 1} & {\bf 2} & - \frac{1}{2} & 1-3x-3y \\
   \psi_R^\ell & {\bf 1} & {\bf 2} & - \frac{1}{2} & 2x-3y \\ \hline
   \psi_L^d & {\bf 3} & {\bf 1} & - \frac{1}{3} & -2x-2y \\
   \psi_R^d & {\bf 3} & {\bf 1} & - \frac{1}{3} & 3x-2y-1 \\ \hline
   H_U & {\bf 1} & {\bf 2} & + \frac{1}{2} & -2x +3y\\
   H_D & {\bf 1} & {\bf 2} & + \frac{1}{2} & 3x+3y-1 \\
   H_{\nu'} & {\bf 1} & {\bf 2} & + \frac{1}{2} & 2-7x+3y \\
   \Phi & {\bf 1} & {\bf 1} & 0 & +2 \\ \hline
  \end{array}
$$
  \end{center}
  \caption{Charge assignment for $U(1)_{10+x\bar{5}+yY}$. \label{charge2}}
\end{table}

\begin{table}
 \begin{center}
$$
  \begin{array}{c|ccc|c} \hline
    & SU(3)_c & SU(2)_W & U(1)_Y & U(1)'_{10+x\bar{5}+yY} \\ \hline
   q_L & {\bf 3} & {\bf 2} & \frac{1}{6} & x + y\\
   u_R & {\bf 3} & {\bf 1} & \frac{2}{3} & -x + 4y\\
   d_R & {\bf 3} & {\bf 1} & - \frac{1}{3} & 3x-2y-1 \\ \hline
   \ell_L & {\bf 1} & {\bf 2} & - \frac{1}{2} & 1-3x-3y \\
   \nu_R & {\bf 1} & {\bf 1} & 0 & -1 \\
   e_R & {\bf 1} & {\bf 1} & -1 & -x-6y \\ \hline
   \nu'_R & {\bf 1} & {\bf 1} & 0 & 1-5x \\  \hline
   \psi_L^\ell & {\bf 1} & {\bf 2} & - \frac{1}{2} & 2x-3y-1 \\
   \psi_R^\ell & {\bf 1} & {\bf 2} & - \frac{1}{2} & 2x-3y \\ \hline
   \psi_L^d & {\bf 3} & {\bf 1} & - \frac{1}{3} & -2x-2y \\
   \psi_R^d & {\bf 3} & {\bf 1} & - \frac{1}{3} & 1-2x-2y \\ \hline
   H_U & {\bf 1} & {\bf 2} & + \frac{1}{2} & -2x+3y \\
   H_D & {\bf 1} & {\bf 2} & + \frac{1}{2} & 1-2x+3y \\
   H_{\nu} & {\bf 1} & {\bf 2} & + \frac{1}{2} & 3x+3y-2 \\
   \Phi & {\bf 1} & {\bf 1} & 0 & +2 \\ \hline
  \end{array}
$$
  \end{center}
  \caption{Charge assignment for $U(1)'_{10+x\bar{5}+yY}$. \label{charge3}}
\end{table}

Let us consider another extension represented by
$U(1)_{10+x\bar{5}}$ in Ref.~\cite{Carena:2004xs}.
The charge assignments are shown in Table~\ref{charge2} and \ref{charge3}.
We here added the current of $U(1)_Y$, which does not change
the charge of $\nu_R$.
We thus call it a $U(1)_{10+x\bar{5}+yY}^{'}$ model.
In this extension, we introduce extra vector-like fermions $\psi_{L,R}^\ell$
and $\psi_{L,R}^d$ with respect to the SM charges.
Also, there are two kinds of the right-handed neutrinos, which have
no SM charges.
We always impose $Q_{\nu_R} = -1$ and then the other right-handed neutrino 
has a different $U(1)$ charge.
In passing, the charges of $d_R$ and $\ell_L$ in Table~\ref{charge3}
are just the replacements by those of $\psi_R^d$ and $\psi_L^\ell$
in Table~\ref{charge2}, respectively.
Depending on them, the charges of the Higgs doublets are determined. 
In Table~\ref{charge2}, the SM Yukawa couplings are
\begin{equation}
 -{\cal L}_y = y_u \bar{q}_L u_R \tilde{H}_U + y_d \bar{q}_L d_R H_D
  + y_\ell \bar{\ell}_L e_R H_D + y_\nu \bar{\ell}_L \nu_R \tilde{H}_U
  + y_{\nu'} \bar{\ell}_L \nu'_R \tilde{H}_{\nu'},
\end{equation}
while they are 
\begin{equation}
 -{\cal L}_y = y_u \bar{q}_L u_R \tilde{H}_U + y_d \bar{q}_L d_R H_D
  + y_\ell \bar{\ell}_L e_R H_D + y_\nu \bar{\ell}_L \nu_R \tilde{H}_\nu
  + y_{\nu'} \bar{\ell}_L \nu'_R \tilde{H}_U,
\end{equation}
in Table~\ref{charge3}.

The coefficients $a$ of $\beta_{g_{1'}}$ are
\begin{equation}
  a = \left( 80 x^2 - 40 x + 120 y^2 + 8 \right) \, N_g 
    + \frac{4}{3} N_\Phi + \frac{2}{3} (2x-3y)^2 N_{H_U} 
    + \frac{2}{3} (3x+3y-1)^2 N_{H_D}
    + \frac{2}{3} (7x-3y-2)^2 N_{H_{\nu'}} ,
\end{equation}
for $U(1)_{10+x\bar{5}+yY}$ in Table~\ref{charge2}, 
and
\begin{equation}
  a = \left( 80 x^2 - 40 x + 120 y^2 + 8 \right) \, N_g 
    + \frac{4}{3} N_\Phi + \frac{2}{3} (2x-3y)^2 N_{H_U} 
    + \frac{2}{3} (2x-3y-1)^2 N_{H_D}
    + \frac{2}{3} (3x+3y-2)^2 N_{H_{\nu}} ,
\end{equation}
for $U(1)'_{10+x\bar{5}+yY}$ in Table~\ref{charge3}.
In both cases, we find
\begin{equation}
  a  > \left( 80 x^2 - 40 x + 120 y^2 + 8 \right) \, N_g 
    + \frac{4}{3} N_\Phi ,
\end{equation}
by ignoring the contributions of the Higgs doublets,
and thus, for $N_g=3$ and $N_\Phi=1$, the lower bound of $a$ is
$a > 31/3$ at $x = 1/4$ and $y=0$. 
Therefore we obtain
\begin{equation}
 K > 1.111, \quad 1.179, \quad 1.155 \quad
 \mbox{for} \quad N_\nu=1,2,3, 
\end{equation}
respectively.
We conclude that the flatland scenario does not work in the models 
in Table~\ref{charge2} and \ref{charge3}. 
\subsection{$U(1)_{xd-u+yY}$}

\begin{table}
 \begin{center}
$$
  \begin{array}{c|ccc|c} \hline
    & SU(3)_c & SU(2)_W & U(1)_Y & U(1)_{xd-u+yY} \\ \hline
   q_L & {\bf 3} & {\bf 2} & \frac{1}{6} & y \\
   u_R & {\bf 3} & {\bf 1} & \frac{2}{3} & -1+4y \\
   d_R & {\bf 3} & {\bf 1} & - \frac{1}{3} & x-2y \\ \hline
   \ell_L & {\bf 1} & {\bf 2} & - \frac{1}{2} & 1-x-3y \\
   \nu_R & {\bf 1} & {\bf 1} & 0 & -1 \\
   e_R & {\bf 1} & {\bf 1} & -1 & 1-6y \\ \hline
   \psi_L^\ell & {\bf 1} & {\bf 2} & - \frac{1}{2} & \frac{1}{5}+x-3y \\
   \psi_R^\ell & {\bf 1} & {\bf 2} & - \frac{1}{2} & \frac{6}{5}-3y \\ \hline
   \psi_L^d & {\bf 3} & {\bf 1} & - \frac{1}{3} & -\frac{1}{5}-2y \\
   \psi_R^d & {\bf 3} & {\bf 1} & - \frac{1}{3} & \frac{4}{5} - x -2y \\ \hline
   H_U & {\bf 1} & {\bf 2} & + \frac{1}{2} & -1 +3y\\
   H_D & {\bf 1} & {\bf 2} & + \frac{1}{2} & -x+3y \\
   H_\nu & {\bf 1} & {\bf 2} & + \frac{1}{2} & x+3y-2 \\
   \Phi & {\bf 1} & {\bf 1} & 0 & +2 \\ \hline
  \end{array}
$$
  \end{center}
  \caption{Charge assignment for $U(1)_{xd-u+yY}$. \label{charge4}}
\end{table}

We next consider a model represented by
$U(1)_{d-xu}$ in Ref.~\cite{Carena:2004xs}.
We added the current of $U(1)_Y$ and thus call it a $U(1)_{xd-u+yY}$ model.
For the charge assignment, see Table~\ref{charge4}.
Note that
we rescaled the charge so as to get $Q_{\nu_R} = -1$.
We corrected the typos of the charge assignments $\psi_{L,R}^{\ell,d}$
in Ref.~\cite{Carena:2004xs}.

From  Table~\ref{charge4}, the following SM Yukawa couplings are written;
\begin{equation}
 -{\cal L}_y = y_u \bar{q}_L u_R \tilde{H}_U + y_d \bar{q}_L d_R H_D
  + y_\ell \bar{\ell}_L e_R H_D + y_\nu \bar{\ell}_L \nu_R \tilde{H}_\nu\,.
\end{equation}

The coefficient $a$ of $\beta_{g_{1'}}$ is
\begin{equation}
  a = \left( \frac{20}{3} x^2 - \frac{16}{3} x
             + 120 y^2 - 48 y + 8 \right) \, N_g 
    + \frac{4}{3} N_\Phi + \frac{2}{3} (3y-1)^2 N_{H_U} 
    + \frac{2}{3} (x-3y)^2 N_{H_D}
    + \frac{2}{3} (x+3y-2)^2 N_{H_{\nu}} \, .
\end{equation}
For $N_g=3$ and $N_\Phi=N_{H_U}=N_{H_D}=N_{H_\nu}=1$,
the minimum of $a$ is 
$a_{\rm min} = 713/84$, when $x=x_{\rm min} = 7/16$, and
$y=y_{\rm min} = 13/63$.
We then find
\begin{equation}
 K_{\rm min} = 0.98579, \quad 1.04559, \quad 1.02446 \quad
 \mbox{for} \quad N_\nu=1,2,3, 
\end{equation}
respectively.
Even for $N_g=3$, this model can work, if we take $N_\nu=1$.

We summarize the results in the subsections A, B, and C in Table~\ref{tab2}.

\begin{table}
$$
  \begin{array}{c|c|l} \hline
    \mbox{model} & \mbox{coeff. of } \beta_{g_{1'}} & 
    \hfil \mbox{lower bounds of $K$} \\ \hline
    U(1)_{xq - \tau^3_R} &
    a = (80 x^2 -32 x + \frac{16}{3}) \, N_g 
      + \frac{4}{3} N_\Phi + \frac{2}{3} (3x-1)^2 N_H &
    \begin{array}{c|l}
      N_\nu = 1 & K \geqq 0.9415 \\
      N_\nu = 2 & K \geqq 0.9986 \\
      N_\nu = 3 & K \geqq 0.9785 \\
    \end{array} \\ \hline
    U(1)_{10+x\bar{5}+yY} &
    a > ( 80 x^2 - 40 x + 120 y^2 + 8 ) \, N_g 
    + \frac{4}{3} N_\Phi &
    \begin{array}{c|l}
      N_\nu = 1 & K > 1.111 \\
      N_\nu = 2 & K > 1.179 \\
      N_\nu = 3 & K > 1.155 \\
    \end{array} \\ \hline
    U(1)_{xd-u+yY} &
    \begin{array}{l}
    a = ( \frac{20}{3} x^2 - \frac{16}{3} x + 120 y^2 - 48 y + 8 ) \, N_g 
    + \frac{4}{3} N_\Phi \\ \quad + \frac{2}{3} (3y-1)^2 N_{H_U} 
    + \frac{2}{3} (x-3y)^2 N_{H_D}
    + \frac{2}{3} (x+3y-2)^2 N_{H_{\nu}}
    \end{array}
   &
    \begin{array}{c|l}
      N_\nu = 1 & K \geqq 0.9858 \\
      N_\nu = 2 & K \geqq 1.046 \\
      N_\nu = 3 & K \geqq 1.024 \\
    \end{array} \\ \hline
  \end{array}
$$
  \caption{Coefficients of the $\beta$ function of $g_{1'}$ and 
   lower bounds of $K$ for various $U(1)$ extension 
   models~\cite{Carena:2004xs}. 
   We fixed to $N_g=3$ and $N_\Phi=N_H=1$
   (or $N_\Phi=N_{H_U}=N_{H_D}=N_{H_\nu}=1$).
   The flatland  scenario is possible only when $K<1$.}
  \label{tab2}
\end{table}
\subsection{Minimal vector-like $U(1)'$}

\begin{table}
 \begin{center}
$$
  \begin{array}{c|ccccc} \hline
   & SU(3)_c & SU(2)_W & U(1)_Y & U(1)' & Z_2 \\ \hline
   \nu_R & {\bf 1} & {\bf 1} & 0 & -1 & + 1 \\
   N_R & {\bf 1} & {\bf 1} & 0 & +1 & -1 \\ \hline
   \psi_L & {\bf 1} & {\bf 1} & x & y & -1 \\
   \psi_R & {\bf 1} & {\bf 1} & x & y & -1 \\ \hline
   \Phi & {\bf 1} & {\bf 1} & 0 & +2 & + 1 \\ \hline
  \end{array}
$$
  \end{center}
  \caption{Charge assignment for the minimal vector-like $U(1)'$ model.
   The SM fermions and the Higgs doublet have no $U(1)'$ charge and 
   do even parity of the $Z_2$ symmetry. \label{simplest}}
\end{table}

In the previous subsections, we studied several chiral $U(1)'$ models.
Simpler is a vector-like $U(1)'$ model.
Let us consider such a model including the right-handed neutrinos 
$\nu_R$.

Unlike the previous chiral $U(1)'$ extension, 
we may divide the model into the SM and CW sectors.
In this case, only the CW sector has the $U(1)'$ charge.
In order to mediate the SSB of the CW sector to the EWSB in the SM sector, 
we also need to introduce some fields having both charges of
the SM gauge interactions and the extra $U(1)'$.
This scenario can be realized by the minimal set of 
the required matter contents.
Also, we note that this setup is similar to 
the dark $Z'$ models~\cite{dark-Z}.  

For the anomaly cancellation from $\nu_R$,
we introduce the right-handed fermion $N_R$ with 
the opposite $U(1)'$ charge to $\nu_R$ for each generation.
The Majorana Yukawa couplings among $\Phi$, $\nu_R$ and $N_R$ are
\begin{equation}
 - {\cal L}_M = 
  Y_M^{ij} \overline{\nu_{Ri}^c} \Phi \nu_{Rj}
  + Y_N^{ij} \overline{N_{Ri}} \Phi N_{Rj}^c
  + \mbox{(h.c.)} \, . 
\end{equation}
The Majorana mass term $\overline{\nu_R^c} N_R$ is dangerous, 
because it potentially yields a big correction to the mass term of $\Phi$
at the 1-loop level.
We thus assign different $Z_2$ parity to $\nu_R$ and $N_R$ in order to forbid
this Majorana mass term.
As an intermediary between the CW and SM sectors, 
we also introduce vector-like fermions $\psi_{L,R}$ with both charges of 
$U(1)_Y$ and $U(1)'$.
These fermions causes the nonzero value of the quartic Higgs mixing coupling
in low energy through the gauge mixing effects.
Although the vector-like mass term for $\psi_{L,R}$ breaks 
the classical conformality, it does not contribute to the mass terms 
for $H$ and $\Phi$ at the 1-loop level, because there is no interaction 
among $\psi_{L,R}$, $H$ and $\Phi$.
In this sense, the theory is still ``natural''.
We show the charge assignment in Table~\ref{simplest}.

In this model, the coefficient $a$ of $\beta_{g_{1'}}$ is
\begin{equation}
  a = \frac{4}{3} N_g + \frac{4}{3} N_\Phi + \frac{4y^2}{3} N_\psi,
\end{equation}
where $N_\psi$ is the number of $\psi_{L,R}$, and $N_g=3$.
For simplicity, we may take $Y_M^{ij}$ and $Y_N^{ij}$ as
diagonal matrices, 
$Y_{M (N)}^{ij} = \diag (y_{M (N)},\cdots,y_{M (N)}, 0,\cdots,0)$,
as was done in the previous subsection, 
and further simplify them to $y_M=y_N$. In this case,
\begin{equation}
  b=4+2N_{\nu_R}+2N_{N_R}, \quad c= 6, \quad d = 16 (N_{\nu_R} + N_{N_R}), \quad 
  f=96 ,
\end{equation}
where $N_{\nu_R}$ and $N_{N_R}$ denote the number of $\nu_R$ and $N_R$
having the relevant Majorana Yukawa couplings, respectively.
The full set of the RGEs are shown in Appendix~\ref{app-vec}.

In this model, we find
\begin{equation}
  K = \frac{\frac{2}{3} N_g + \frac{2}{3} N_\Phi + \frac{2y^2}{3}N_\psi + 3}
           {2+N_{\nu_R}+N_{N_R}}
      \sqrt{\frac{N_{\nu_R} + N_{N_R}}{6}} \, .
\end{equation}
For simplicity, taking $N_g=3$, $N_\Phi=1$, $N_\psi=1$, $N_{\nu_R} = N_{N_R}$, 
and $y=1$, we obtain 
\begin{equation}
  K = 0.914, \; 0.862, \; 0.792, \quad \mbox{for} \quad 
  N_{\nu_R} = N_{N_R} = 1, \; 2, \; 3 \, .
\end{equation}
In this way, the flatland scenario is easily realized 
in the minimal vector-like model.
\section{$\rho$-parameter}
\label{sec3}

In the previous section, we introduced (multiple) Higgs doublet(s) 
with extra $U(1)$ charges.
This is, however, very dangerous, because
the $\rho$-parameter deviates from unity at the tree level.

Let us revisit a formula of the $\rho$-parameter~\cite{Langacker:2008yv}.

In general, we introduce multiple Higgs doublets $H_k$ 
with $k=1,2,\cdots,N_H$.
The hypercharge of $H_k$ is commonly $Y_H=1/2$ by definition and
the extra $U(1)$ charges are $Y'_k$.
The VEVs of $H_k$ are represented by $v_k$.
The $\Phi$ for the extra $U(1)$ breaking does not have the hypercharge.
The covariant derivative is then
\begin{equation}
  D_\mu = \partial_\mu - i g_2 \frac{\tau^i}{2} W_\mu^i
 - i Y (g_Y B_\mu + \gtld B'_\mu) - i g_{1'} Y' B'_{\mu} , 
\end{equation}
where the gauge mixing $\gtld$ appears and 
$Y'$ denotes the charge of $U(1)'$.
After diagonalizing the mass matrix of the neutral gauge boson, 
we obtain the masses of $Z$ and $Z'$ ($M_Z < M_{Z'}$).
The tree level $\rho$-parameter is defined by
$\rho_0 = M_W^2 /c_W^2 M_Z^2$,
and hence we obtain a simple formula for 
the deviation of the $\rho$-parameter from unity as follows:
\begin{eqnarray}
\delta \rho \equiv \rho_0 -1 =
 \frac{\tan \theta_{ZZ'} \sum_k (\gtld + 2 Y'_k g_{1'}) \frac{v_k^2}{v^2}}
      {\frac{g_2}{c_W} - \tan \theta_{ZZ'} 
        \sum_k (\gtld + 2 Y'_k g_{1'}) \frac{v_k^2}{v^2}},
 \label{drho}
\end{eqnarray}
with the $Z$--$Z'$ mixing $\theta_{ZZ'}$,
\begin{equation}
  \tan \theta_{ZZ'} = \frac{c_W}{g_2} \frac{M_0^2}{M_{Z'}^2 - M_0^2}
  \sum_k (\gtld + 2 Y'_k g_{1'}) \frac{v_k^2}{v^2} ,
\end{equation}
where we used the SM formula of the $Z$ mass,
$M_0^2 \equiv g_2^2 v^2/(4 c_W^2)$, and
$c_W = \cos \theta_W = e/g_Y$.
Since the $Z$ mass $M_Z$ is always smaller than the SM formula $M_0$,
we generally find $\delta \rho > 0$.

The numerator of Eq.~(\ref{drho}) should be small.
Otherwise, $\delta\rho$ becomes terribly large.
We then obtain approximately,
\begin{eqnarray}
  \delta \rho \simeq 
   \frac{v^2}{4M_{Z'}^2}
    \left(\sum_k (\gtld + 2 Y'_k g_{1'}) \frac{v_k^2}{v^2}\right)^2 \, .
\end{eqnarray}
Compared with the experimental bound
$\rho_0 = 1.0004^{+0.0003}_{- 0.0004}$~\cite{pdg},
we find
\begin{equation}
  \frac{v}{M_{Z'}} \left|\sum_k (\gtld + 2 Y'_k g_{1'}) \frac{v_k^2}{v^2}\right|
  \lesssim 0.05 \, .
  \label{rho-bound}
\end{equation}
In terms of $\theta_{ZZ'}$ with $M_{Z'} \sim 1$~TeV,
the above inequality corresponds to $|\theta_{ZZ'}| \lesssim 10^{-3}$,
which agrees with the conventionally quoted bound~\cite{Langacker:2008yv}.  
Since the bound by the particle data group~\cite{pdg} includes 
higher order corrections, one might not take the number of 
the inequality (\ref{rho-bound}) at face value.
At least an order of magnitude of the $Z$--$Z'$ mixing effect
is limited by (\ref{rho-bound}).

In order to construct realistic models, 
we need to survey other constraints from the precision measurements, 
the bounds of the direct searches at Tevatron and LHC, 
etc.~\cite{Cacciapaglia:2006pk,Erler:2009jh,Salvioni:2009mt}.  
These constraints depend on details of the charge assignments of 
the SM fermions, however. 
Such model-dependent analyses will be performed elsewhere.
Last but not least, it is noticeable that 
the $Z$--$Z'$ mixing angle $\theta_{ZZ'}$ is observable at the LHC and/or 
at the future ILC~\cite{Weihs:2011wp,Choi:2013qra}. 

\section{Realizations of flatland scenario}
\label{sec4}

As concrete realizations of the flatland scenario, 
we take the chiral $U(1)_{xq-\tau_R^3}$ and 
vector-like $U(1)'$ models in Sec.~\ref{sec2}. 
The potential $V$ for $H$ and $\Phi$ is 
\begin{equation}
V = \lambda_H |H|^4 +\lambda_\Phi |\Phi|^4 + \lambda_{\rm mix} |H|^2 |\Phi|^2 ,
 \label{V}
\end{equation}
where we assumed the classically conformality for the scalar fields.

In the flatland scenario, 
we impose vanishing of the scalar potential at a UV scale $\Lambda$,
\begin{equation}
 \lambda_H (\Lambda)=0, \quad \lambda_\Phi (\Lambda)=0, \quad
 \lambda_{\mathrm{mix}} (\Lambda)=0.
 \label{flatboundary}
\end{equation}
We also set $ \gtld (\Lambda)=0$ 
by constructing a model with no $U(1)$ kinetic mixing 
at the high energy scale $\Lambda$. 

For appropriate parameters investigated below, 
the CWM occurs in the $U(1)'$ sector of $\Phi$
and thereby $\Phi$ acquires the VEV, $\VEV{\Phi} = v_\Phi/\sqrt{2}$,
~\cite{Iso:2009ss,Iso:2012jn}.
Then the EWSB takes place if 
the $H$--$\Phi$ mixing term $\lambda_{\mathrm{mix}} |H|^2 |\Phi|^2$
is negative;
\begin{equation}
  v_H^2 = \frac{- \lambda_{\mathrm{mix}}}{2\lambda_H} \, v_\Phi^2 ,
  \label{vH}
\end{equation}
where $\VEV{H} = (0, \; v_H/\sqrt{2})^T$. 
The Higgs mass $m_H$ is approximately given 
by $m_H^2 = 2\lambda_H v_H^2$, because the mixing 
between $H$ and $\Phi$ is tiny.
In \cite{Iso:2012jn}, we have shown that such a small and 
negative scalar mixing is radiatively generated through 
the gauge kinetic mixing of $U(1)_{B-L}$ and $U(1)_Y$.

Starting from the flat potential (\ref{flatboundary}) at the UV scale,
 running couplings are obtained numerically by solving the RGEs, 
and the value of $v_H$ in Eq.~(\ref{vH}) is predicted in terms of them.
The RG flows are controlled by,  besides the SM parameters, 
the $U(1)'$ gauge coupling and the Majorana Yukawa couplings at $\Lambda$.
Given these two parameters at $\Lambda$, the symmetry breaking scales of 
$\Phi$ and $H$ are determined.
In order to set $v_H = 246$~GeV, we must adjust one of the two parameters
in accordance with the other. 
Hence there is only {\it one free parameter} in the model.
In particular, the CW relation
\begin{equation}
   \frac{m_\phi^2}{M_{Z'}^2}
 + N_\nu \frac{g_{1'}^2}{\pi^2}\frac{M_{\nu_R}^4}{M_{Z'}^4}
 \simeq \frac{3g_{1'}^2}{2\pi^2} 
 \label{mzp-mphi-mnuR}
\end{equation}
must hold, where we used
\begin{equation}
 m_\phi^2 \simeq -4 \lambda_\Phi v_\Phi^2,  \quad
  M_{Z'} \simeq 2 g_{1'} v_\Phi, \quad M_{\nu_R} \simeq \sqrt{2}\, y_M v_\Phi \,.
\end{equation}
Note that the CW relation comes from the relation between 
the running scalar coupling $\lambda_\Phi$ and 
the $\beta$ function $\beta_{\lambda_\Phi}$; 
$\beta_{\lambda_\Phi} \simeq -4 \lambda_\Phi$ at the CW scale 
$\mu=v_\Phi$ (see e.g., eq.(4) in \cite{Iso:2009ss}). 

In the following analysis, we take the UV scale at
$\Lambda=1/\sqrt{8\pi G} = 2.435 \times 10^{18}$~GeV.
Also, we fix the Higgs mass, $m_h=126.8$~GeV
and the $\overline{MS}$ mass of the top quark\footnote{
This is consistent with the indirect prediction, 
$\overline{m_t}=167.5^{+8.9}_{-7.3}$~GeV~\cite{pdg}, while
the converted value to the pole mass~\cite{Melnikov:2000qh} 
is rather small, compared to
the directly obtained value at the Tevatron/LHC.} 
$\overline{m_t} \simeq 160$~GeV so as to realize $\lambda_H(\Lambda)=0$.

\subsection{$U(1)_{xq-\tau_R^3}$}

Let us analyze the minimal $U(1)_{xq-\tau_R^3}$ model with $x_{\rm min}=25/123$,
where the value of $K$ is minimized.
The $U(1)_{xq-\tau_R^3}$ charge for $H$ causes $\lambda_{\rm mix} \ne 0$ 
in the IR scale.
This is potentially dangerous, because 
the $\rho$-parameter deviates from unity at the tree level.
We also note that in the minimal $U(1)_{xq-\tau_R^3}$ model,
the $\beta$ function for the gauge mixing term $\gtld$ is vanishing. 
Because the charge of $U(1)_{xq-\tau_R^3}$ is given by $Q=6xY-\tau_R^3$,
we find the derivative of Eq.~(\ref{coeff-a}) with respect to $x$ as
\begin{equation}
  \frac{\partial a}{\partial x} =
  \frac{4}{3} \sum_f Q_f \frac{\partial Q_f}{\partial x} + 
  \frac{2}{3} \sum_f Q_s \frac{\partial Q_s}{\partial x} = 
  12 \bigg[\, \frac{2}{3} \sum_f Q_f Y_f + \frac{1}{3} \sum_s Q_s Y_s\,\bigg]
  = 12 \, a_{\rm mix}, 
\end{equation}
where $a_{\rm mix}$ is the coefficient of the $\beta$ function 
for $\gtld$, which is not proportional to $\gtld$
(see  Appendix~\ref{app-u1p}).
If we take $x=x_{\rm min}$, 
$\partial a/\partial x = 0$ holds and $a_{\rm mix}=0$ is satisfied.
Then if we set the gauge mixing term as $\gtld(\Lambda) = 0$ 
at the UV scale $\Lambda$, 
it continues to be zero at any energy scale; $\gtld(\mu) \equiv 0$.
This property reduces the labor of the numerical analysis.

\begin{figure}[t]
  \begin{center}
  \resizebox{0.3\textheight}{!}
            {\includegraphics{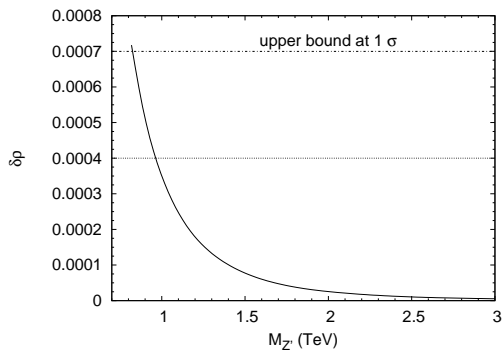}}
  \end{center}
  \caption{$\delta \rho$ for the minimal $U(1)_{xq-\tau_R^3}$ model
   with $x=25/123$.
  The dotted and dash-dotted lines correspond to 
  the central value and the upper bound at $1 \sigma$, respectively.
  The results are almost the same for $N_\nu=1,2,3$, where 
  $N_\nu$ represents the number of the right-handed neutrinos having
  the relevant Majorana Yukawa couplings.
  \label{drho-u1p}}
\end{figure}

We show the values of $\delta \rho$ in Fig.~\ref{drho-u1p}.
Allowing $1\sigma$ deviation, we find the lower bound for 
the $Z'$ mass as $M_{Z'} \gtrsim 0.8$~TeV.
The results are almost independent of the number of $\nu_R$
having the relevant Majorana Yukawa couplings.

\begin{figure}[t]
  \begin{center}
  \resizebox{0.3\textheight}{!}
            {\includegraphics{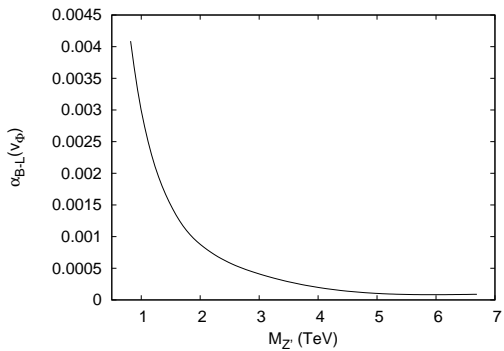}}
  \end{center}
  \caption{$M_{Z'}$ v.s. $\alpha_{B-L}(v_\Phi)$ 
   for the minimal $U(1)_{xq-\tau_R^3}$ model with $x=25/123$.
  The results are almost the same for $N_\nu=1,2,3$.
  \label{Mzp_vs_alp_u1p}}
\end{figure}

We also depict the values of $\alpha_{B-L} = g_{B-L}^2/(4\pi)$
at $\mu=v_\Phi$ in Fig.~\ref{Mzp_vs_alp_u1p}.
We here used the notation $\gBL$ instead of $g_{1'}$ in order to 
compare the results with 
the previous ones~\cite{Iso:2009ss,Iso:2009nw,Iso:2012jn,Hashimoto:2013hta}.
In comparison with the pure $B-L$ model shown in Fig.~3 
in Ref.~\cite{Hashimoto:2013hta}, 
the obtained gauge coupling is so small
and thereby $\delta \rho$ is reduced from the naively expected one.

\begin{figure}[t]
  \begin{center}
  \resizebox{0.3\textheight}{!}
            {\includegraphics{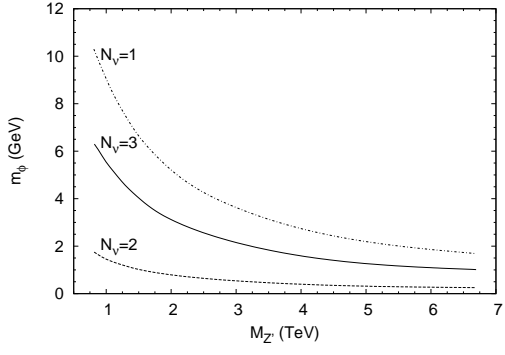}}
  \end{center}
  \caption{$M_{Z'}$ v.s. $m_\phi$ for the minimal $U(1)_{xq-\tau_R^3}$ model
   with $x=25/123$.
  $N_\nu$ denotes the number of the right-handed neutrinos having
  the relevant Majorana Yukawa couplings.
  \label{Mzp_vs_mphi_u1p}}
\end{figure}

We further show the extra Higgs mass $m_\phi$ in Fig.~\ref{Mzp_vs_mphi_u1p}.
Although the values of $m_\phi$ depend on $N_\nu$, 
we find roughly $m_\phi \sim {\cal O}(\mbox{GeV})$.
Compared with the pure $B-L$ model shown in Fig.~2 
in Ref.~\cite{Hashimoto:2013hta}, $m_\phi$ is also suppressed 
because of the very tiny $\beta$ function $|\beta_{\lambda_\Phi}| \ll 1$.
The mass of the right-handed neutrinos $M_{\nu_R}$ 
can be easily estimated from the CW relation~(\ref{mzp-mphi-mnuR}),
$M_{\nu_R} \simeq \sqrt[4]{\frac{3}{2N_\nu}} M_{Z'}$, 
i.e., the left-hand side of (\ref{mzp-mphi-mnuR}) is saturated by $M_{\nu_R}$.

\subsection{Minimal vector-like $U(1)'$}

Let us analyze the minimal vector-like $U(1)'$ model.
For simplicity, we take $N_g=3$, $N_\Phi = N_\psi = N_{\nu_R} = N_{N_R}=1$,
and $x=y=1$. 

\begin{figure}[t]
  \begin{center}
  \resizebox{0.3\textheight}{!}
            {\includegraphics{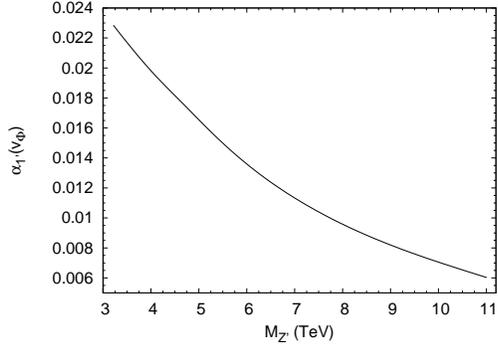}}
  \end{center}
  \caption{$M_{Z'}$ v.s. $\alpha_{1'}(v_\Phi)$ 
  for the minimal vector-like $U(1)'$ model.
  We took $N_\Phi=N_\psi=N_{\nu_R} = N_{N_R}=1$ and $x=y=1$.
  \label{Mzp_vs_alp_u1v}}
\end{figure}

We depict the results for $\alpha_{1'} = g_{1'}^2/(4\pi)$
at $\mu=v_\Phi$ in Fig.~\ref{Mzp_vs_alp_u1v}.
The values and behaviors are similar to those in 
the pure $B-L$ model shown in Fig.~3 
in Ref.~\cite{Hashimoto:2013hta}. 
In this model, the deviation of the $\rho$-parameter comes only from 
$\gtld \ne 0$ in low energy, so that it is tiny, 
$\delta \rho \lesssim 10^{-6}$. 

\begin{figure}[t]
  \begin{center}
  \resizebox{0.3\textheight}{!}
            {\includegraphics{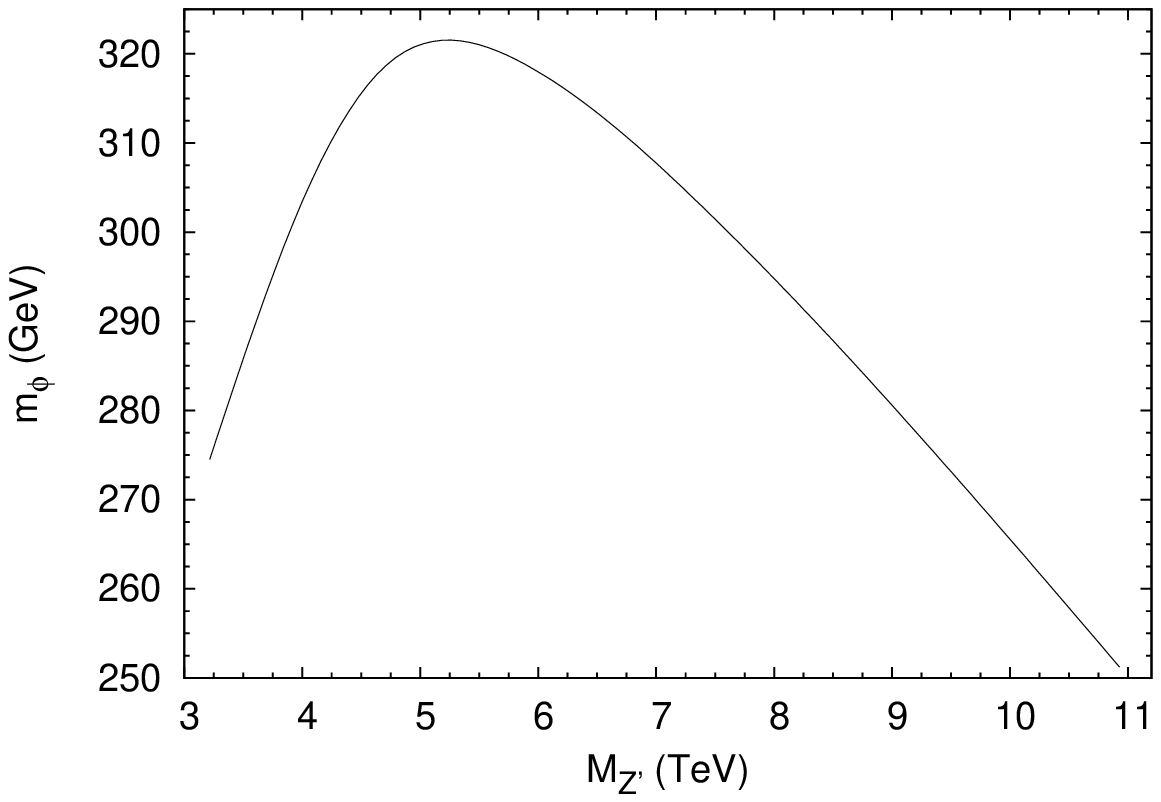}}
  \end{center}
  \caption{$M_{Z'}$ v.s. $m_\phi$ for the minimal vector-like $U(1)'$ model.
  We took $N_\Phi=N_\psi=N_{\nu_R} = N_{N_R}=1$ and $x=y=1$.
  \label{Mzp_vs_mphi_u1v}}
\end{figure}

We also show the extra Higgs mass $m_\phi$ in Fig.~\ref{Mzp_vs_mphi_u1v}.
The values of $m_\phi$ are, say, $m_\phi \sim {\cal O}(\mbox{100 GeV})$,
and much larger than those of $U(1)_{xq-\tau_R^3}$.
This is because $K$ is closer to one, and consequently 
$|\beta_{\lambda_\Phi}| \ll 1$,
in the minimal $U(1)_{xq-\tau_R^3}$ model.

\section{Summary}
\label{summary}

We investigated several models with a $U(1)$ extension and checked
whether  the condition $K < 1$
for the flatland scenario is satisfied. 
As shown in Ref.~\cite{Hashimoto:2013hta}, 
the pure $B-L$ models with $N_g=1,2$ and $N_\nu=1$ satisfy $K < 1$, 
while the familiar $B-L$ model with $N_g=3$ does not.
We found that  twisted versions of the $B-L$ model, 
$U(1)_{xq-\tau_R^3}$, and $U(1)_{xd-u+yY}$ models satisfy 
the condition even for $N_g=3$. 
We also proposed a minimal vector-like $U(1)'$ model with $K < 1$. 

In particular we explicitly calculated the behavior of 
the running coupling constants in the minimal $U(1)_{xq-\tau_R^3}$ model 
and in the minimal vector-like $U(1)'$ model, 
and confirmed the realizations of the flatland scenario. 
Although the $\rho$-parameter deviates from unity at the tree level
in the $U(1)_{xq-\tau_R^3}$ model, the deviation is small for 
$M_{Z'} \gtrsim 0.8$~TeV because of the very small gauge coupling.
For the minimal vector-like $U(1)'$ model, 
since the deviation $\delta \rho \ne 0$ appears essentially from 
the one-loop corrections, it is suppressed as
$\delta \rho \lesssim 10^{-6}$; the model is not strongly constrained 
by the $\rho$-parameter.
The setup of the minimal vector-like $U(1)'$ model is similar 
to the semi-invisible $Z'$ model~\cite{dark-Z}.  
The $Z_2$-odd fermion $N_R$ in the model can be 
a candidate of the dark matter.
Also, the $Z$--$Z'$ mixing effect can be explored at the LHC and/or 
at the future ILC~\cite{Weihs:2011wp,Choi:2013qra}. 
Further investigations will be performed elsewhere.

\appendix

\section{RGEs for $U(1)_{xq-\tau_R^3}$}
\label{app-u1p}

We show the full set of the RGEs for the $U(1)_{xq-\tau_R^3}$ model.

The RGEs for the SM gauge couplings are 
\begin{equation}
(16 \pi^2) \mu \frac{\partial }{\partial \mu} g_i = c_i g_i^3,  
\end{equation}
with
\begin{equation}
  c_Y = \frac{41}{6}, \quad
  c_2 = - \frac{19}{6}, \quad
  c_3 = -7,
\end{equation}
The RGEs for $U(1)'$ are
\begin{eqnarray}
  (16 \pi^2) \mu \frac{\partial }{\partial \mu} \gBL &=&
  \gBL \bigg[\, a \, \gBLsq + 2 a_{\rm mix} \, \gBL \gtld
  + c_Y \gtldsq\,\bigg], \\
  (16 \pi^2) \mu \frac{\partial }{\partial \mu} \gtld &=&
  \gtld \bigg[\, c_Y \, (\gtldsq + 2 g_Y^2) + a \, \gBLsq\,\bigg]
  + 2 a_{\rm mix} \, \gBL (\gtldsq + g_Y^2),
\end{eqnarray}
with
\begin{eqnarray}
  a &=& \left( 80 x^2 -32 x + \frac{16}{3} \right) \, N_g 
     +  \frac{4}{3} N_\Phi + \frac{2}{3} (3x-1)^2 N_H, \\
 a_{\rm mix} &=& \left(\frac{40}{3}x - \frac{8}{3}\right) N_g
     +  \left(x-\frac{1}{3}\right) N_H \, .
\end{eqnarray}

For the Yukawa couplings, we find
\begin{equation}
  (16 \pi^2) \mu \frac{\partial }{\partial \mu} y_t =
  y_t \bigg[\, \frac{9}{2} y_t^2
  - \left(8g_3^2 + \frac{9}{4}g_2^2+\frac{17}{12} (g_Y^2 + \gtldsq)
    + 3 (17x^2 - 8x +1) \gBLsq + (17x-4) \gBL\,\gtld\right)\,\bigg],
\end{equation}
and
\begin{equation}
  (16 \pi^2) \mu \frac{\partial }{\partial \mu} y_M =
  y_M \bigg[\, (4 + 2 N_\nu) y_M^2 - 6 \gBLsq\,\bigg],
\end{equation}
where $Y_{M}^{ij} = \diag (y_{M},\cdots,y_M, 0,\cdots, 0)$,
$\tr[(Y_{M}^{ij})^2] = N_\nu y_M^2$, etc. with 
$N_\nu$ being the number of the large Majorana couplings.
The RGEs for the Higgs quartic couplings are
\begin{eqnarray}
  (16 \pi^2) \mu \frac{\partial \lambda_H}{\partial \mu} &=&
   24 \lambda_H^2 + \lambda_{\rm mix}^2
  + \frac{3}{8}\bigg(2g_2^4 + 
     \big\{g_2^2+g_Y^2+(\gtld+2(3x-1)\gBL)^2\big\}^2\bigg)
  \nonumber \\ &&
  - 3 \lambda_H \big\{3 g_2^2 + g_Y^2 + (\gtld + 2(3x-1) \gBL)^2\big\}
  - 6 y_t^4 + 12 \lambda_H y_t^2, \\
  (16 \pi^2) \mu \frac{\partial \lambda_\Phi}{\partial \mu} &=&
   20\lambda_\Phi^2 + 2\lambda_{\rm mix}^2 - 16 N_\nu y_M^4
   + 8 N_\nu \lambda_\Phi y_M^2
  + 96 \gBLquad - 48 \lambda_\Phi \gBLsq, \\
  (8 \pi^2) \mu \frac{\partial \lambda_{\rm mix}}{\partial \mu} &=&
   \lambda_{\rm mix} \bigg[\,
   6\lambda_H + 4 \lambda_\Phi + 2\lambda_{\rm mix} + 3 y_t^2 + 2N_\nu y_M^2
  -\frac{3}{4}\big\{3g_2^2 + g_Y^2 + (\gtld+2(3x-1)\gBL)^2\big\}
  - 12 \gBLsq\,\bigg] 
  \nonumber \\ &&
  + 6 (\gtld+2(3x-1)\gBL)^2 \, \gBLsq \, .
\end{eqnarray}
 
The minimal $U(1)_{xq-\tau_R^3}$ corresponds to
$x=x_{\rm min}=\frac{25}{123}$, which yields
$\min (a)  = \frac{964}{123}$ and minimizes $K$.
In this case, we also find $a_{\rm mix}=0$, so that $\gtld(\mu) \equiv 0$
in all energy scale.
This choice is allowed, because
$\lambda_{\rm mix}(\Lambda)=0$ does not mean $\lambda_{\rm mix}(\mu) \equiv 0$
at any $\mu$, owing to the $\gBLquad$ term in 
$\frac{\partial \lambda_{\rm mix}}{\partial \ln \mu}$.
Even in this case, the values of $\delta \rho$ are kept 
within the experimental bounds, if $M_{Z'} \gtrsim 0.8$~TeV.
 
\section{RGEs for the minimal vector-like $U(1)'$ model}
\label{app-vec}

We show the full set of the RGEs for the minimal vector-like $U(1)'$ model.

The RGEs for the SM gauge couplings are 
\begin{equation}
(16 \pi^2) \mu \frac{\partial }{\partial \mu} g_i = c_i g_i^3,  
\end{equation}
with
\begin{equation}
  c_Y = \frac{41}{6} + \frac{4x^2}{3} N_\psi, \quad
  c_2 = - \frac{19}{6}, \quad
  c_3 = -7,
\end{equation}
The RGEs of $g_{1'}$ and $\gtld$ are
\begin{eqnarray}
  (16 \pi^2) \mu \frac{\partial }{\partial \mu} g_{1'} &=&
  g_{1'} \bigg[\, a \, g_{1'}^2 + 2a_{\rm mix} \, g_{1'} \gtld
  + c_Y \gtldsq\,\bigg], \\
  (16 \pi^2) \mu \frac{\partial }{\partial \mu} \gtld &=&
  \gtld \bigg[\, c_Y \, (\gtldsq + 2 g_Y^2) + a \, g_{1'}^2\,\bigg]
  + 2 a_{\rm mix}\, g_{1'} (\gtldsq + g_Y^2),
\end{eqnarray}
with
\begin{equation}
  a = \frac{4}{3} N_g + \frac{4}{3} N_\Phi + \frac{4y^2}{3} N_\psi, \qquad
  a_{\rm mix} = \frac{4xy}{3} N_\psi \, .
\end{equation}

The RGE for the top Yukawa coupling is
\begin{equation}
  (16 \pi^2) \mu \frac{\partial }{\partial \mu} y_t =
  y_t \bigg[\, \frac{9}{2} y_t^2
  - \left(8g_3^2 + \frac{9}{4}g_2^2+\frac{17}{12} (g_Y^2 + \gtldsq)
    \right)\,\bigg],
\end{equation}
and also those for the Majorana Yukawa couplings are
\begin{equation}
  (16 \pi^2) \mu \frac{\partial }{\partial \mu} y_M =
  y_M \bigg[\, (4 + 2N_{\nu_R}) y_M^2 + 2 N_{N_R} y_N^2 - 6 g_{1'}^2\,\bigg],
\end{equation}
and
\begin{equation}
  (16 \pi^2) \mu \frac{\partial }{\partial \mu} y_N =
  y_N \bigg[\, (4 + 2N_{N_R}) y_N^2 + 2 N_{\nu_R} y_M^2 - 6 g_{1'}^2\,\bigg],
\end{equation}
where we took $Y_{M (N)}^{ij} = \diag (y_{M (N)},\cdots,0,\cdots)$,
$\tr[(Y_{M (N)}^{ij})^2] = N_{\nu_R (N_R)} y_{M (N)}^2$, etc. with 
$N_{\nu_R (N_R)}$ being the number of the large Majorana couplings.

The RGEs for the scalar quartic couplings are
\begin{eqnarray}
  (16 \pi^2) \mu \frac{\partial \lambda_H}{\partial \mu} &=&
   24 \lambda_H^2 + \lambda_{\rm mix}^2
  + \frac{3}{8}\bigg(2g_2^4 + (g_2^2+g_Y^2+\gtldsq)^2\bigg)
  - 6 y_t^4 + 12 \lambda_H y_t^2
  - 3 \lambda_H (3 g_2^2 + g_Y^2+\gtldsq), \\
  (16 \pi^2) \mu \frac{\partial \lambda_\Phi}{\partial \mu} &=&
   20\lambda_\Phi^2 + 2\lambda_{\rm mix}^2 - 16 (N_{\nu_R} y_M^4 + N_{N_R} y_N^4)
   + 8 (N_{\nu_R} y_M^2 + N_{N_R} y_N^2) \lambda_\Phi 
  + 96 g_{1'}^4 - 48 \lambda_\Phi g_{1'}^2, \\
  (8 \pi^2) \mu \frac{\partial \lambda_{\rm mix}}{\partial \mu} &=&
   \lambda_{\rm mix} \bigg[\,
   6\lambda_H + 4 \lambda_\Phi + 2\lambda_{\rm mix} + 3 y_t^2 + 2 N_{\nu_R} y_M^2
  + 2 N_{N_R} y_N^2 -\frac{3}{4}(3g_2^2 + g_Y^2 + \gtldsq) - 12 g_{1'}^2\,\bigg]
  \nonumber \\ &&
  + 6 g_{1'}^2 \, \gtldsq \, .
\end{eqnarray}

\end{document}